\RequirePackage{fixltx2e}
\documentclass[a4paper]{isp}
\usepackage[american]{babel}
\usepackage{graphicx}
\usepackage{amsmath,amsthm,mathtools}
\usepackage{url}
\usepackage[capitalise,nameinlink]{cleveref}
\usepackage{xcolor}
\def\aap{A\&A\,  }

\def\aj{AJ  }
\def\apj{ApJ\,  }
\def\apss{Astrophysics and Space Science  }

\def\mnras{MNRAS\,  }

\def\rmxaa{Revista Mexicana de Astronomia y Astrofisica}

\def\sun{\hbox{$\odot$}}
\def\s2{$\rm{[SII]}$}
\begin{document}
\pdfgentounicode=1
\title
{
Models for  velocity decrease in HH34 
}
\author{Lorenzo Zaninetti}
\institute{
Physics Department,
 via P.Giuria 1, I-10125 Turin,Italy \\
 \email{zaninetti@ph.unito.it}
}

\maketitle

\begin {abstract}
The conservation of the energy flux  in turbulent jets
that propagate in the interstellar medium (ISM)
allows us to deduce the  law of motion
when an  inverse power law decrease of density is considered.
The back-reaction that is caused by the radiative losses for the trajectory 
is evaluated.
The velocity dependence  of the jet with time/space 
is applied to the jet of HH34, for which the 
astronomical data of velocity versus time/space are available.
The introduction of  precession and constant velocity
for the central star allows us to build a curved trajectory
for the superjet connected with HH34. 
The bow shock that is visible in the superjet is explained 
in the framework of the theory of the image 
in the case of an optically thin layer.
\end{abstract}

\section{Introduction}

The equation of motion plays a relevant role in our understanding of  
the physics of the Herbig-Haro objects (HH), after  
\cite{Herbig1950,Haro1952}.
A common example is to evaluate the 
velocity of the jet in HH34  as 300 $km/s$, see \cite{Raga2019},
without paying  attention to its spatial or temporal
evolution.
A precise evaluation of the evolution of the jet's velocity with time 
in HH34 has been done, for example, by 
\cite{Raga2012}.
It is therefore possible to speak of {\it proper motions
of young stellar outflows}, 
see \cite{Raga2013,Raga2014,Guzman2016,Raga2017}.

The first set of theoretical efforts exclude 
the magnetic field:
\cite{Cabrit2000} 
have modeled the  slowing down of the HH 34 superjet 
 as a result of the jet's interaction  
with the surrounding environment,
\cite{Lopez2001} 
have shown that a velocity profile in the jet beam is 
        required to explain
        the observed acceleration in the position-velocity diagram of
        the HH jet,
\cite{Viti2003}
         found some  constraints on the  the physical and chemical
        parameters of the clump ahead of HHs 
and
\cite{Raga2011}
reviewed some  important understanding of outflows from young stars.

The second set of    theoretical efforts include  
the magnetic field:
\cite{Kwon2010} 
analysed the  HH 1-2 region in the L1641 molecular 
cloud 
and found  a straight magnetic field  of about  130 micro-Gauss,
\cite{Lee2014}
analysed HH 211  and found 
field lines of the magnetic field with different orientations, 
\cite{Lee2016} analysed HH 111 and
found evidence for magnetic braking.

These theoretical efforts  
to understand HH objects 
leave a series of questions
unanswered or partially answered, as follows:
\begin {itemize}
\item Is it possible to find  a law of motion 
      for turbulent jets in the presence of a medium 
      with a density that  decreases as a power law?
\item Is it possible to introduce the  back reaction
      into the equation of motion for turbulent jets 
      to model the radiative losses?
\item Can we model the bending of the super-jet 
      connected with HH34?  
\item Can we explain the bow shock  visible in 
      in HH34 with the theory of the image?
\end{itemize}
To answer these questions,
this paper 
reviews in Section \ref{sec_preliminaries}
the velocity observations of HH34 at a 9 yr time interval,
Section \ref{section_simple} 
analyses  two simple models as given by 
the  Stoke's and Newton's
laws of resistance,
Section  \ref{secenergy} 
applies the conservation of the energy    
flux in a turbulent jet to find an equation of motion,
Section \ref{secextended} models the extended 
region of HH34 , the so called "superjet", 
and Section \ref{sec_image}  reports some analytical and 
numerical algorithms that allow us to build the image of HH34.

\section{Preliminaries}
\label{sec_preliminaries}

The velocity evolution of the HH34 jet has 
recently been analysed  in \s2, (672 nm),
 frames
and   Table I in  \cite{Raga2012}  reports the Cartesian coordinates,
the velocities,  
and  the dynamical time 
for  18 knots in  9 years of observations.
To start with time, $t$, equal to zero,
we fitted the velocity versus distance with the following power law
\begin{equation} 
v(x;x_0,v_0) = v_0\times (x/x_0)^\alpha 
\label{equationvx}
\quad ,
\end{equation}
where $v$ and $x$ are  the velocity  and the length of the jet,
$v_0$  is the velocity at $x=x_0$
and $\alpha$ with  its relative error 
is a parameter   to be  found with a
fitting procedure.
The integration of this equation gives  the time 
as a function of the position, $x$,
as given
by the fit
\begin{equation}
t=
-{\frac {{{\it x}}^{-\alpha+1}{x_0}^{\alpha}-x_0}{ \left( \alpha-1
 \right) v_0}}
\quad ,
\label{eqntime}
\end{equation}
where $x_0$ is the position at $t=0$.
The  fitted trajectory, distance  versus time,  is  
\begin{equation}
x(t;x_0,v_0)
=
{{\rm e}^{{\frac {\alpha\,\ln  \left( x_0 \right) -\ln  \left( -t v_0\,
\alpha+t v_0+ x_0 \right) }{\alpha-1}}}}
\quad , 
\end{equation}
and the fitted velocity as function of time is  
\begin{equation}
v(t;x_0,v_0)
=
v_0\, \left( {\frac {1}{x_0}{{\rm e}^{{\frac {\alpha\,\ln  \left( x_0
 \right) -\ln  \left( -t \left( \alpha-1 \right) v_0+x_0 \right) }{
\alpha-1}}}}} \right) ^{\alpha}
\quad .
\end{equation}
The adopted physical units are $pc$ for length  and $year$ for time,
and the useful conversion for the velocity is 
$1\frac{pc}{year} = 979682.5397 \frac{km}{s}$.

\begin{figure}
\includegraphics[width=7cm,angle=-90]{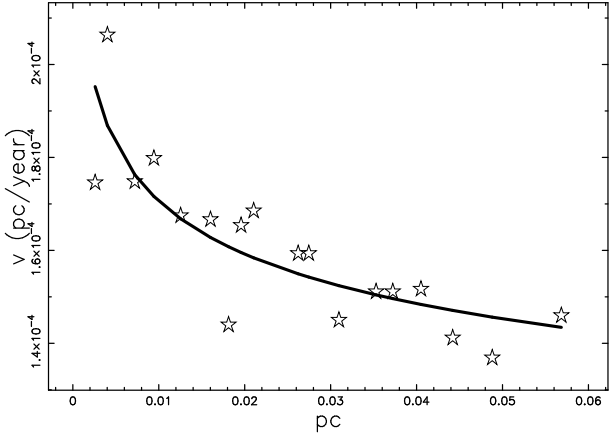}
\caption
{
Observational points of velocity in $pc/yr$ versus distance in $pc$
(empty circles) and
best fit as given by equation (\ref{equationvx}) (full line).
}
\label{vxfit}
\end{figure}

The fit of equation (\ref{equationvx})  
when $x$ is expressed in $pc$
gives
\begin{equation} 
v(x) =  
0.000107\,{x}^{- 0.0998 \pm  0.01618}
\frac{pc}{yr}  
\label{vxfitnumer}
\quad ,
\end{equation}
from which we can conclude that the velocity decreases 
with increasing distance, see  Figure \ref{vxfit}.

The time is derived  from equation (\ref{eqntime}) 
and 
Table \ref{tablehh34} reports the basic parameters of HH34.
This time is more continuous in respect to the dynamical time
reported in  column 6 of  Table I in  \cite{Raga2012}.
\begin{table}[ht!]
\caption
{
Numerical values 
for the physical parameters of HH34 
}
\label{tablehh34}
\begin{center}
\begin{tabular}{|c|c|c|c|}
\hline
knot  &  x(pc) & v(km/s) & time (yr) \\
\hline
 1&  0.002594&  171.01    &  0.\\
 2&  0.004021&  202.15    &   7.484\\
 3&  0.007214&  171.23    &  25.131\\
 4&  0.009434&  176.13    &  37.903\\
 5&  0.012545&  164.04    &  56.303\\
 6&  0.016017&  163.24    &  77.381\\
 7&  0.018125&  141.03    &  90.415\\
 8&  0.019580&  162.00    &  99.502\\
 9&  0.021046&  165.07    &  108.71\\
 10&  0.02622&  156.08    &  141.76\\
 11&  0.02745&  156.08    &  149.75\\
 12&  0.03096&  142.03    &  172.63\\
 13&  0.03528&  148.03    &  201.20\\
 14&  0.03723&  148.03    &  214.17\\
 15&  0.04050&  148.569855&  236.15\\
 16&  0.04420&  138.293167&  261.17\\
 17&  0.04880&  134.082062&  292.60\\
 18&  0.05684&  143.003494&  348.25\\
\hline
\end{tabular}
\end{center}
\end{table}

\section{Two simple models}
\label{section_simple}
When a  jet   moves through  the interstellar medium (ISM),
a retarding  drag
force, $F_{drag}$, is applied. 
If $v$ is the instantaneous velocity, then
the simplest model assumes 
\begin {equation}
F_{drag} \propto  v ^n
\quad  ,
\end {equation}
where   $n$ is  an integer.
Here, the case  of  $n=1$ and $n=2$ is considered.
In classical mechanics,  $n=1$ is referred to  as Stoke's
law of resistance and   $n=2$ is referred to as Newton's
law of resistance.

\subsection{Stoke's  behaviour}

\label{sec_stokes}
\begin{figure}
\includegraphics[width=7cm,angle=-90]{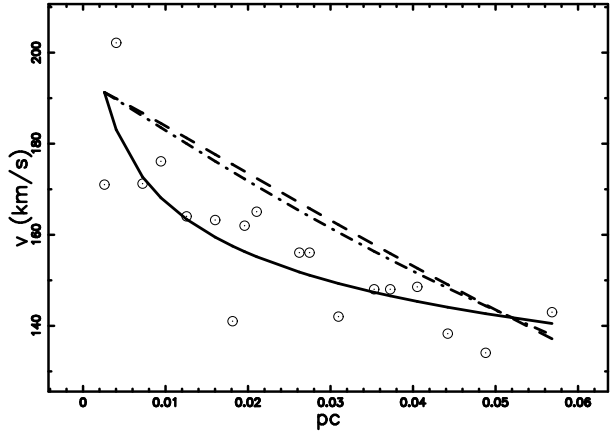}
\caption
{
Observational points of velocity in $pc/yr$ versus distance in $pc$
(empty circles),
best fit as given by equation  (\ref{equationvx})    (full line),
Stokes behaviour as given by equation (\ref{vxstokes})(dashed line)
and 
Newton behaviour as given by equation (\ref{vxnewton})
(dot-dash-dot-dash).
}
\label{vxkms}
\end{figure}

The equation of motion is given by
\begin{equation}
\frac {dv(t)}{dt} = -B  v(t)
\quad .
\end  {equation}
The  velocity as function of time   is
\begin{equation}
v   =v_0 e ^{ -B t}
\label{sv}
\quad ,
\end  {equation}
where $v_0$  is the initial  velocity.
The distance at     time $t$  is
\begin{equation}
x  = s \left( t \right) =x_{{0}}-{\frac {v_{{0}}{{\rm e}^{-Bt}}}{B}}+{
\frac {v_{{0}}}{B}}
\quad  .
\label{first}
\end  {equation}
The time as function of distance is obtained by
the inversion of this equation
\begin{equation}
t= \frac{1}{B} -\ln  \left( -{\frac {xB-Bx_{{0}}-v_{{0}}}{v_{{0}}}} \right)
\quad .
\end{equation}
The velocity as a function of space is
\begin{equation}
v(x;x_0,v_0,B)= -xB+Bx_{{0}}+v_{{0}}
\label{vxstokes}
\quad .
\end{equation}
The numerical value of $B$ is
\begin{equation}
B=-{\frac {v_{{0}}-v_{{1}}}{x_{{0}}-x_{{1}}}}
\quad ,
\end{equation}
where $v_1$ is the velocity at point $x_1$;
the data of Table \ref{tablehh34} gives 
B= 0.0009549, see Figure~\ref{vxkms}.

\subsection{Newton's  behaviour}

\label{sec_newton}
The equation of motion is  
\begin{equation}
\frac {dv(t)}{dt} = -  A v(t)^2
\quad .
\end  {equation}
The  velocity as function of time   is
\begin{equation}
v   =v \left( t \right) ={\frac {v_{{0}}}{Atv_{{0}}+1}}
\label{vt}
\quad ,
\end  {equation}
where $v_0$  is the initial  velocity.
The distance at    time $t$  is
\begin{equation}
x  = s \left( t \right) ={\frac {\ln  \left( Atv_{{0}}+1 \right) }{A}}+x_{{0
}}
\quad  .
\end  {equation}
The time as function of distance is obtained by
the inversion of the above equation
\begin{equation}
t= {\frac {{{\rm e}^{xA-Ax_{{0}}}}-1}{Av_{{0}}}}
\quad .
\end{equation}
The velocity as function of the distance is
\begin{equation}
v(x;x_0,v_0,A) =
{\frac {v_{{0}}}{{{\rm e}^{xA-Ax_{{0}}}}}}
\label{vxnewton}
\quad .
\end{equation}

The numerical value of $A$ is
\begin{equation}
A=-{\frac {1}{x_{{0}}-x_{{1}}}\ln  \left( {\frac {v_{{0}}}{v_{{1}}}}
 \right) }
\quad ,
\end{equation}
where $v_1$ is the velocity at point $x_1$;
the data of Table \ref{tablehh34} gives 
A= 5.68381834.

\section{Energy flux conservation}

\label{secenergy}

The conservation of the energy    flux in a turbulent jet
requires a perpendicular section to the motion along the
Cartesian $x$-axis, $A$
\begin {equation}
A(r)=\pi~r^2
\end{equation}
where $r$ is the radius of the jet.
Section  $A$ at  position $x_0$  is
\begin {equation}
A(x_0)=\pi ( x_0   \tan ( \frac{\alpha}{2}))^2
\end{equation}
where   $\alpha$  is the opening angle and
$x_0$ is the initial position on the $x$-axis.
At position $x$, we have
\begin {equation}
A(x)=\pi ( x   \tan ( \frac{\alpha}{2}))^2
\quad .
\end{equation}
The conservation  of energy flux states that
\begin{equation}
\frac{1}{2} \rho(x_0)  v_0^3   A(x_0)  =
\frac{1}{2} \rho(x  )   v(x)^3 A(x)[B
\label{conservazioneenergy}
\end {equation}
where $v(x)$ is the velocity at  position $x$ and
$v_0(x_0)$   is the velocity at  position $x_0$,
see Formula A28 in \cite{deyoung}.
More details can be found in \cite{Zaninetti2016e,Zaninetti2018b}.
The density  is  assumed to decrease as
a power law
\begin{equation}
\rho = \rho_0  (\frac{x_0}{x})^{\delta}
\label{profpower}
\end{equation}
where  $\rho_0$ is the density at  $x=x_0$
and $\delta$ a positive parameter.
The differential equation that  models the
energy  flux is
\begin{equation}
\frac{1}{2}\, \left( {\frac {x_{{0}}}{x}} \right) ^{\delta} \left( {\frac 
{\rm d}{{\rm d}t}}x \left( t \right)  \right) ^{3}{x}^{2}-\frac{1}{2}\,{v_{{0}
}}^{3}{x_{{0}}}^{2}
=0
\quad  .
\end{equation}
The velocity as a function of the position, $x$,
\begin{equation}
v(x) =
\frac
{
\sqrt [3]{{x_{{0}}}^{2} \left(  \left( {\frac {x_{{0}}}{x}} \right) ^{
\delta} \right) ^{2}x}v_{{0}}
}
{
\left( {\frac {x_{{0}}}{x}} \right) ^{\delta}x
}
\quad .
\label{velocitypower}
\end{equation}

\begin{figure}
\includegraphics[width=7cm,angle=-90]{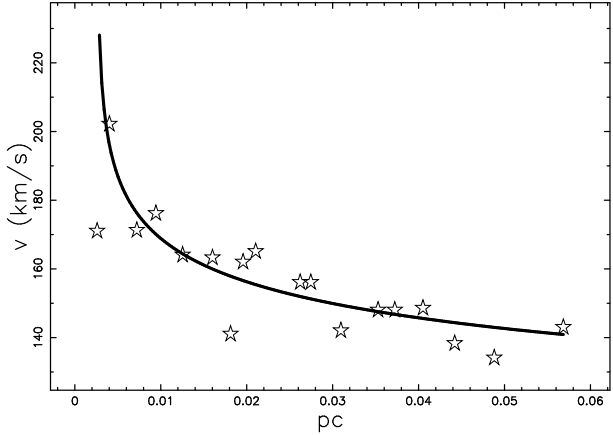}
\caption
{
Observational points of velocity in $pc/yr$ versus distance in $pc$
(empty stars).
The theoretical  fit is given by equation 
(\ref{velocitypower}) (full line)
with parameters  $x_0= 0.00259 \,pc$ ,$v_0 =191.27\,\frac{km}{s}$
and  $\delta=1.7$.
}
\label{hh34turb}
\end{figure}

Figure \ref{hh34turb} reports the velocity as a function 
of the distance and the observed points.

We now have four models for the velocity as a function of time 
and   Table \ref{chi2hh34} 
reports 
the merit function $\chi^2$, which is evaluated
as
\begin{equation} 
\chi^2 =\sum_{i=1}^N \Big 
[  y_{i,theo}-y_{i,obs}  \Big ]^2
\label{chiquaredef}
\end{equation}
where $y_{i,obs}$   represents
the observed value  at position $i$
and $y_{i,theo}$ the theoretical value at position $i$.
\begin{table}[ht!]
\caption
{
The values of the $\chi^2$ for four models of velocity
of HH34 
}
\label{chi2hh34}
\begin{center}
\begin{tabular}{|c|c|}
\hline
Model  & $\chi^2$  \\
\hline
power law fit (no physics)     &  1479   \\
Stoke's  behaviour             &  3813   \\
Newton's  behaviour            &  3317   \\
turbulent jet                  &  2373   \\
\hline
\end{tabular}
\end{center}
\end{table}
A careful analysis  of Table \ref{chi2hh34}
allows us to  conclude that the turbulent jet 
performs  better in respect to the Stokes's and Newton's behavior.

The trajectory , i.e. the distance as function of the time,
\begin{equation}
\label{xtturb}
x(t;r_0,v_o,\delta)
=
x_{{0}}{{\rm e}^{{\frac {1}{\delta-5} \left( 3\,\ln  \left( 3 \right) 
-3\,\ln  \left( 5-\delta \right) -\ln  \left( {\frac {{t}^{3}{v_{{0}}}
^{3}}{{x_{{0}}}^{3}}} \right)  \right) }}}
\quad ,
\end{equation}
and the velocity as function of time
\begin{eqnarray}
v(t;r_0,v_0,\delta)
=\frac{1}{x_0}
{3}^{{\frac {-2\,\delta+1}{\delta-5}}} \left( 5-\delta \right) ^{{
\frac {2\,\delta-1}{\delta-5}}}\sqrt [3]{{x_{{0}}}^{3} \left(  \left( 
{\frac {{t}^{3}{v_{{0}}}^{3}}{{x_{{0}}}^{3}}} \right) ^{ \left( \delta
-5 \right) ^{-1}} \right) ^{2\,\delta} \left( {\frac {{t}^{3}{v_{{0}}}
^{3}}{{x_{{0}}}^{3}}} \right) ^{- \left( \delta-5 \right) ^{-1}}}
\times  
\nonumber \\
v_{{0
}} \left( {{\rm e}^{{\frac {1}{\delta-5} \left( -3\,\ln  \left( 3
 \right) +3\,\ln  \left( 5-\delta \right) +\ln  \left( {\frac {{t}^{3}
{v_{{0}}}^{3}}{{x_{{0}}}^{3}}} \right)  \right) }}} \right) ^{-\delta}
{{\rm e}^{{\frac {1}{\delta-5} \left( -3\,\ln  \left( 3 \right) +3\,
\ln  \left( 5-\delta \right) +\ln  \left( {\frac {{t}^{3}{v_{{0}}}^{3}
}{{x_{{0}}}^{3}}} \right)  \right) }}}
\end{eqnarray}

\begin{figure}
\includegraphics[width=7cm,angle=-90]{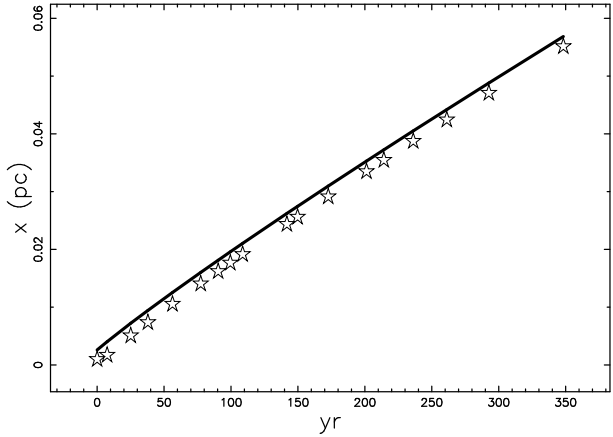}
\caption
{
Observational points of distance  in $pc$ versus time  in $years$
(empty stars).
The theoretical  curve  is given by equation 
(\ref{xtturb}) (full line)
with the same parameters 
as in Figure \ref{hh34turb}.
}
\label{hh34xt}
\end{figure}

Figure \ref{hh34xt} reports the trajectory  as a function 
of time and of the observed points.

The rate of mass flow at the point $x$, $\dot {m}(x)$, is
\begin{equation}
\dot {m}(x;v,alpha) =
\rho   v(x)  \pi ( x   \tan ( \frac{\alpha}{2}))^2
\end{equation}
and the astrophysical version is
\begin{equation}
\dot {m}(x;x_0,v_{0,km/s},M_{\sun},\alpha) =
7.92529\,10^{-8}\,{\it n }\,{x}^{4/3-2/3\,\delta} \left( 
\tan \left( \alpha/2 \right)  \right) ^{2}{x_{{0}}}^{2/3+2/3\,\delta}v
_{{0,km/s}}  \frac{ M_{\sun}}{yr}
\end{equation}
where 
$\alpha$ is the opening angle in rad,  
$x$ and $x_0$ are expressed in pc,
$n $ is the number density of protons   
at  $x=x_0$ expressed  in
particles~cm$^{-3}$,
$M_{\sun}$ is the solar mass
and $v_{0,km/s}$ is the initial velocity at point $x_0$  
expressed in $km/s$.
This rate of mass flow as function of the
distance $x$ increases  when $\delta < 2 $, is
constant when $\delta = 2 $,
and decreases when $\delta  >2 $.

\subsection{The back reaction}

Let us suppose that the radiative losses
are  proportional to the   flux of energy 
\begin{equation}
- \epsilon 
\frac{1}{2} \rho(x  )   v(x)^3 A(x)
\quad ,
\end{equation}
where $\epsilon$ is a constant that is  thought 
to be $\ll 1$.
By inserting in the above equation the considered area ,$A(x)$,
and the power law density here adopted
the radiative losses are 
\begin{equation}
-\epsilon
\frac{1}{2}  
\rho_{{0}} \left( {\frac {x_{{0}}}{x}} \right) ^{\delta}{v}^{3}\pi\,{x
}^{2} \left( \tan \left(\frac{ \alpha}{2} \right)  \right) ^{2}
\quad .
\end{equation}

By inserting in this equation  the  velocity to first order
as  given by equation~(\ref{velocitypower}), 
the radiative losses, $Q(x;x_0,v_0,\delta,\epsilon)$, are
\begin{equation}
Q(x;x_0,v_0,\delta,\epsilon)= - \epsilon 
\frac{1}{2}
\rho_{{0}}{v_{{0}}}^{3}\pi\,{x_{{0}}}^{2} \left( \tan \left( \alpha/2
 \right)  \right) ^{2}x
\quad ,
\label{lossesclassical}
\end{equation}
The sum of the radiative  losses between $x_0$ and $x$ 
is given by the following integral, $L$,
\begin{equation}
L(x;x_0,v_0,\delta,\epsilon)=\int_{x_0}^x  Q(x;x_0,v_0,delta,\epsilon) dx
=
-\epsilon\,
\frac
{1}
{2}
\rho_{{0}}{v_{{0}}}^{3}\pi\,{x_{{0}}}^{2} \left( \tan
 \left(\frac{ \alpha}{2} \right)  \right) ^{2} \left( x-x_{{0}} \right) 
\quad .
\label{classiclosses}
\end{equation}
The  conservation of the   flux of energy  in the presence  
of  the back-reaction due to the radiative losses
is 
\begin{eqnarray}
\frac{1}{2}
\left( {v_{{0}}}^{3}{x_{{0}}}^{2}\epsilon\,x-{v_{{0}}}^{3}{x_{{0}}}^{
3}\epsilon+ \left( {\frac {x_{{0}}}{x}} \right) ^{\delta}{v}^{3}{x}^{2
} \right) \rho_{{0}}
=
\frac{1}{2}
\rho_{{0}}{v_{{0}}}^{3}{x_{{0}}}^{2}
\end{eqnarray}
The real  solution  of the cubic equation for  the velocity to 
the second order, $v_c(x;\delta,x_0,v_0)$, 
is
\begin{equation}
v_c(x;\delta,x_0,v_0)=
\sqrt [3]{- \left( \epsilon\,x-\epsilon\,x_{{0}}-1 \right) {x_{{0}}}^{
2+2\,\delta}{x}^{4-2\,\delta}}v_{{0}}{x}^{-2+\delta}{x_{{0}}}^{-\delta
}
\quad .
\label{vback}
\end{equation}

\begin{figure}
\includegraphics[width=7cm,angle=-90]{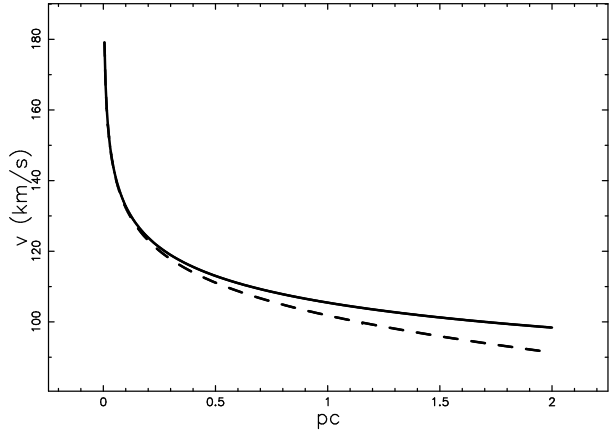}
\caption
{
Velocity to the second order as function of the distance,
see equation (\ref{vback}),
when 
$\epsilon =0$   (full   line) and
$\epsilon =0.1$ (dashed line),
other parameters as in Figure~\ref{hh34turb}.
}
\label{vbackvx}
\end{figure}

Figure \ref{vbackvx} reports the effect of introducing 
the losses on the velocity as function of the distance
for a given value of $\epsilon$,
i.e. the velocity decreases more quickly.

\begin{figure}
\includegraphics[width=7cm,angle=-90]{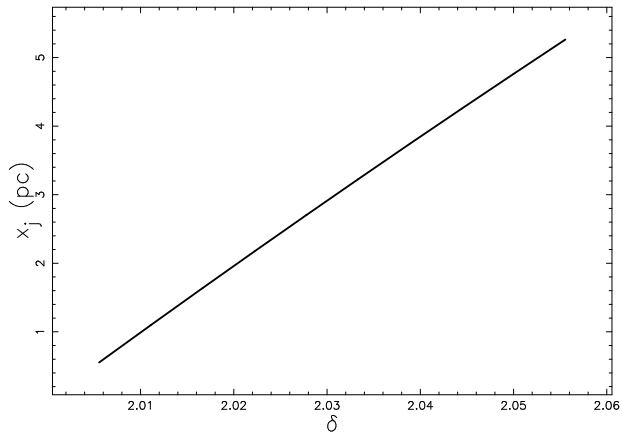}
\caption
{
Length of the  jet as a function  of $\delta$, and the 
other parameters are as in Figure~\ref{hh34turb}.
}
\label{ljet}
\end{figure}

The presence of the  back-reaction  allows us to evaluate  the 
jet's length,  which can be derived from the minimum  
in the corrected velocity to second order as a function of $x$, 
\begin{equation}
\frac{\partial v_c(x;\delta,x_0,v_0)}{\partial x} =0
\quad ,
\end {equation}
which is  
\begin{eqnarray}
\frac 
{
-v_{{0}} \left( \delta\,\epsilon\,x-\delta\,\epsilon\,x_{{0}}-\epsilon
\,x+2\,\epsilon\,x_{{0}}-\delta+2 \right) {x_{{0}}}^{-\delta/3+2/3}{x}
^{-5/3+\delta/3}
}
{
3\, \left( 1+\epsilon\, \left( x_{{0}}-x \right)  \right) ^{2/3}
}
=0
\quad  .
\end{eqnarray}
The solution for $x$ of the above minimum  allows us to derive
the jet's length, $x_j$,
\begin{equation}
x_j = 
{\frac {\delta\,\epsilon\,x_{{0}}-2\,\epsilon\,x_{{0}}+\delta-2}{
\epsilon\, \left( \delta-1 \right) }}
\quad .
\label{jetlength}
\end{equation}
Figure \ref{ljet} reports an example of  the
jet's length  as a function of the parameter $\delta$.
\section{The extended region}

\label{secextended}
To deal with the complex shape of the continuation
of HH34 (e.g. see the new region  HH173 discovered by \cite{Bally1994}),
we should  include the precession of the source and motion
of the host star, following a scheme outlined 
in \cite{Zaninetti2010b}.
The various coordinate systems are
${\bf x}$=$(x,y,z$) , ${\bf x}^{(1)}$=$(x^{(1)},y^{(1)},z^{(1)})$
, $\ldots$ ${\bf x}^{(3)}$=$(x^{(3)},y^{(3)},z^{(3)})$. 
The
vector representing the motion of the jet is 
represented by
the following $1 \times 3$ matrix:
\begin{equation}
G=
\left[ \begin {array}{c} x \left( t \right) \\\noalign{\medskip}0
\\\noalign{\medskip}0\end {array} \right] 
\quad ,
\end{equation}
where the jet motion L(t) is considered along the x axis.

The jet axis, $x$, is inclined at an angle $\Psi_{prec}$
relative to an axis $x^{(1)}$,
 and therefore
the $3 \times 3$ matrix, which
represents a rotation through the z axis,
is given by:
\begin {equation}
F=
 \left[ \begin {array}{ccc} \cos \left( \Psi_{{{\it prec}}} \right) &-
\sin \left( \Psi_{{{\it prec}}} \right) &0\\ \noalign{\medskip}\sin
 \left( \Psi_{{{\it prec}}} \right) &\cos \left( \Psi_{{{\it prec}}}
 \right) &0\\ \noalign{\medskip}0&0&1\end {array} \right] 
\quad .
\end {equation}
The jet is undergoing precession around the
$x^{(1)}$ axis and  $\Omega_{prec}$ is
the angular velocity of precession expressed in
$\mathrm{radians}$ per unit time.
The transformation from the coordinates
${\bf x}^{(1)}$ fixed in the frame of the
precessing jet to the nonprecessing coordinate
${\bf x}^{(2)}$
is represented by the $3 \times 3$ matrix
\begin{equation}
 P=
 \left[ \begin {array}{ccc} 1&0&0\\ \noalign{\medskip}0&\cos \left( 
\Omega_{{{\it prec}}}t \right) &-\sin \left( \Omega_{{{\it prec}}}t
 \right) \\ \noalign{\medskip}0&\sin \left( \Omega_{{{\it prec}}}t
 \right) &\cos \left( \Omega_{{{\it prec}}}t \right) \end {array}
 \right] 
\quad .
\end{equation}

\begin{figure}
\includegraphics[width=7cm,angle=-90]{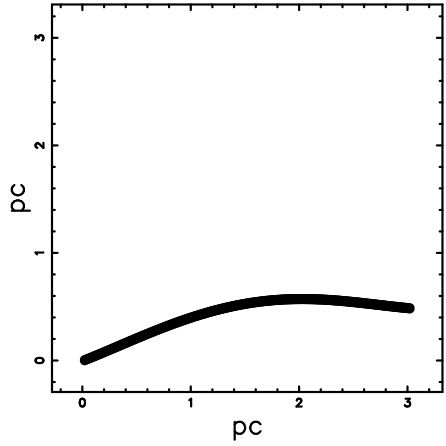}
\caption
{
 Continuous  trajectory of the superjet connected with HH34:
 the three Eulerian angles 
 characterising the point of view are 
   $ \Phi   $= 0 $^{\circ }$
 , $ \Theta $= 0 $^{\circ }$
 and $ \Psi $= 0 $^{\circ }$.
 The precession is characterised by the angle
 $ \Psi_{prec}=10^{\circ}$
 and by the angular velocity
 $ \Omega_{prec} $= 0.00496551674 [$^{\circ}/\mathrm{year}$].
 The star has velocity $v_y=31.107 \frac{Km}{s}$,
 the considered time  is $29000~yr$
and the other parameters are as in Figure~\ref{hh34turb}
}
\label{hh34_front}
\end{figure}

\begin{figure}
\includegraphics[width=7cm,angle=-90]{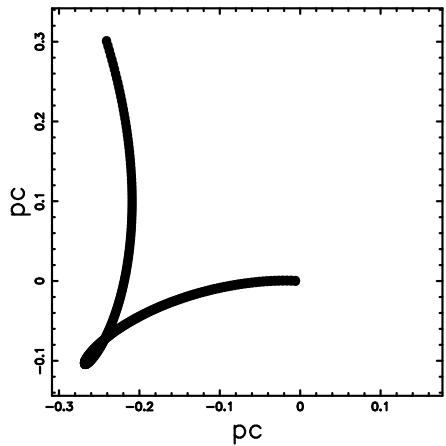}
\caption
{
 Continuous  trajectory of the superjet connected with HH34:
 the three Eulerian angles 
 characterising the point of view are 
   $ \Phi $= 100 $^{\circ }$
 , $ \Theta $= 77 $^{\circ }$
 and $ \Psi $= 135 $^{\circ }$.
The other parameters as in Figure \ref{hh34_front}.
}
\label{hh34_traj}
\end{figure}

The last translation represents
the change of the framework from
$\bf(x^{(2)})$, which is co-moving with
the host star, to a system
$\bf (x^{(3)})$, in comparison to which the
host star is in  a uniform
motion.
The relative motion of the origin of the coordinate system
$(x^{(3)},y^{(3)},z^{(3)})$ is defined by the
Cartesian components of the star velocity $ v_x,v_y,v_z$,
and the required $1 \times 3 $ 
matrix transformation representing this translation is
\begin{equation}
 B=
\left[ \begin {array}{c} v_{{x}}t\\\noalign{\medskip}v_{{y}}t
\\\noalign{\medskip}v_{{z}}t\end {array} \right] 
\label {transla}
\quad .
\end{equation}
On assuming, for the sake 
of simplicity, that $v_x$=0 and $v_z$=0,
the translation matrix becomes
\begin{equation}
 B=
\left[ \begin {array}{c} 0\\\noalign{\medskip}v_{{y}}t
\\\noalign{\medskip}0\end {array} \right] 
\quad .
\label{traslationmatrix}
\end{equation}
 The
final $1 \times 3$ matrix $A$ representing the
``motion law'' can be found
by composing the four matrices already
described;
\begin {eqnarray}
\lefteqn {A = B + ( P \cdot F \cdot G) } \nonumber \\
 =&
\left[ \begin {array}{c} \cos \left( \Psi_{{{\it prec}}} \right) x
 \left( t \right) \\ \noalign{\medskip}v_{{y}}t+\cos \left( \Omega_{{{
\it prec}}}t \right) \sin \left( \Psi_{{{\it prec}}} \right) x \left( 
t \right) \\ \noalign{\medskip}\sin \left( \Omega_{{{\it prec}}}t
 \right) \sin \left( \Psi_{{{\it prec}}} \right) x \left( t \right) 
\end {array} \right] 
\quad .
\label{eqnmatrix}
\end {eqnarray}
The three components of the previous
 $1\times 3$ matrix $A$
represent the jet's motion
along the Cartesian coordinates as given by an  observer
who sees the star moving in a uniform motion.
The  point of view of the observer can be modeled by  
introducing   
the matrix $E$, which represents 
the three Eulerian angles 
$\Theta,\Phi,\Psi$ , see \cite{Goldstein2002}.
A typical  trajectory  is reported in Figure 
\ref{hh34_front}
and a particularised point of view  of the same
trajectory is reported in Figure~\ref{hh34_traj}
in which  a loop is visible.

\section{Image theory}

\label{sec_image}
This section 
summarises the continuum observations of HH34,
reviews the transfer equation with 
a particular attention to 
the case of an optically thin layer,
analyses a simple analytical model for theoretical intensity,
reports
the numerical algorithm that allows us
to build a complex image
and 
introduces the theoretical concept of 
emission from the knots.

\subsection{Observations}

The system of the jet and counter jet of HH34 has been
analysed at 1.5 $\mu m$ and  4.5 $\mu m$, see Figure 3 in  
\cite{Raga2019}.
The intensity is almost constant, 
$I_{1.5} \approx 8\times10^{16} erg\,s^{-1} arcsec^{-2}$
for the first $12^{\prime \prime}$ of the jet
and
$I_{1.5} \approx 3 \times10^{16} erg\,s^{-1} arcsec^{-2}$
for the first $20^{\prime \prime}$ of the counter jet.
At larger distances, the intensity drops monotonically.
At a distance of 414 $pc$ as given by \cite{Raga2012} 
the conversion between physical 
and angular distance is $1\,pc=498.224 ^{\prime \prime}$.
For example, at 1.5 $\mu m$,
the emission is mainly due to the 
${\rm[Fe~II]} 1.64 \mu m$ line.

\subsection{The transfer equation}

For the transfer equation in the presence of emission only
see, for example,  
\cite{rybicki}
 or
\cite{Hjellming1988},
 is
 \begin{equation}
\frac {dI_{\nu}}{ds} =  -k_{\nu} \rho I_{\nu}  + j_{\nu} \rho
\label{equazionetrasfer}
\quad ,
\end {equation}
where  $I_{\nu}$ 
is the specific intensity, 
$s$ is the
line of sight, 
$j_{\nu}$ is the emission coefficient,
$k_{\nu}$ is a mass absorption coefficient,
$\rho$ is the density of mass at position $s$,
and the index $\nu$ denotes the frequency of
emission.
The solution to equation~(\ref{equazionetrasfer})
 is
\begin{equation}
 I_{\nu} (\tau_{\nu}) =
\frac {j_{\nu}}{k_{\nu}} ( 1 - e ^{-\tau_{\nu}(s)} )
\quad  ,
\label{eqn_transfer}
\end {equation}
where $\tau_{\nu}$ is the 
optical depth at frequency $\nu$:
\begin{equation}
d \tau_{\nu} = k_{\nu} \rho ds
\quad.
\end {equation}
We now continue to analyse a  case 
of an optically thin layer
in which $\tau_{\nu}$ is very small
(or $k_{\nu}$ is very small)
and where the density $\rho$ is replaced
by  the concentration $C(s)$
 of the emitting particles:
\begin{equation}
j_{\nu} \rho =K  C(s)
\quad  ,
\end{equation}
where $K$ is a constant.
The intensity is now
\begin{equation}
 I_{\nu} (s) = K
\int_{s_0}^s   C (s\prime) ds\prime \quad  \mbox {optically thin layer},
\label{transport}
\end {equation}
which in the case of constant density, $C$,  is
\begin{equation}
 I_{\nu} (s) = K
C \times (s-s_0)\quad \mbox  {optically thin layer}
\label{transfersimplified}
\quad .
\end{equation}

The increase in brightness
is proportional to the concentration  of particles
integrated along
the line of  sight.

\subsection{Theoretical intensity}

\begin{figure}
\includegraphics[width=7cm,angle=-90]{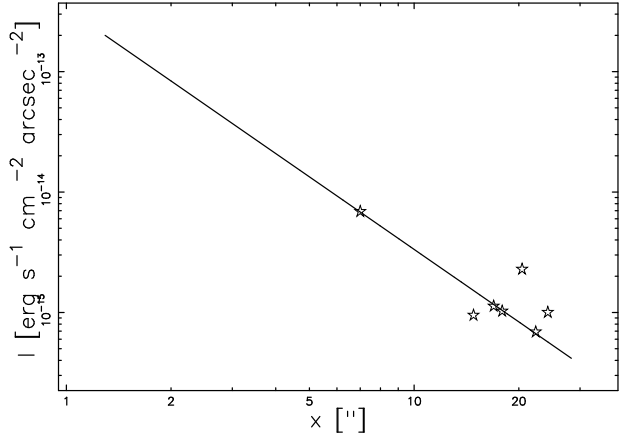}
\caption
{
Observational points of 
intensity at  $4.5\,\mu m$ 
(empty stars) and
theoretical curve  as given 
by equation 
(\ref{equationix}) 
(full line).
when $\epsilon=1/100$, $\rho_0=1$, $\alpha=2.86^{\circ }$ 
and other parameters as in Figure~\ref{hh34turb}.
}
\label{intensity}
\end{figure}

The flux of observed radiation  
along the centre of the jet, $I_c$,  
is  assumed to scale   
as
\begin{equation} 
I_c(x;x_0,v_0,b,\epsilon)\propto \frac{Q(x;x_0,v_0,b,\epsilon)}{x^2}
\label{classicintensity}
\quad  ,
\end{equation}
where  $Q$,
the radiative  losses,  
is   given by equation (\ref{lossesclassical}).
The explicit form of this equation is
\begin{equation}
I_c(x;x_0,v_0,b,\epsilon)=
-\frac{1}{2}\,{\frac { \left( -1+ \left( x-x_{{0}} \right) \epsilon \right) {x
_{{0}}}^{2}{v_{{0}}}^{3}\pi\, \left( \tan \left( \alpha/2 \right) 
 \right) ^{2}\rho_{{0}}}{{x}^{2}}}
\quad .
\label{equationix}
\end{equation}
This relation  connects  the observed 
intensity of radiation with the rate of energy transfer per unit area.
A typical example  for the jet of HH34 at 4.5 $\mu m$
is reported in Figure~\ref{intensity}. 

\subsection{Emission from a cylinder}

\begin{figure}
\includegraphics[width=7cm,angle=-90]{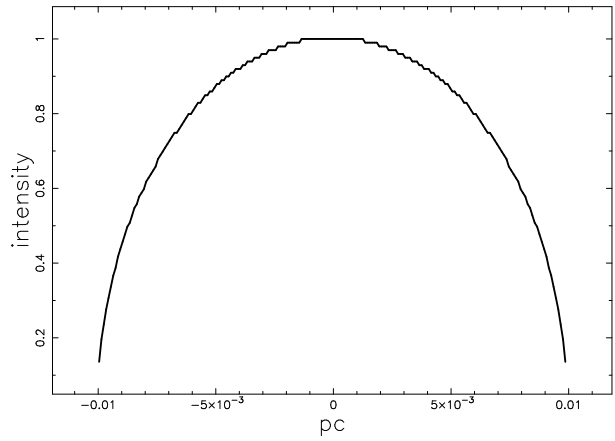}
\caption
{
1D cut of  the intensity, I,  
when  $a=0.01~pc$.
}
\label{cut}
\end{figure}

\begin{figure}
\includegraphics[width=7cm,angle=-90]{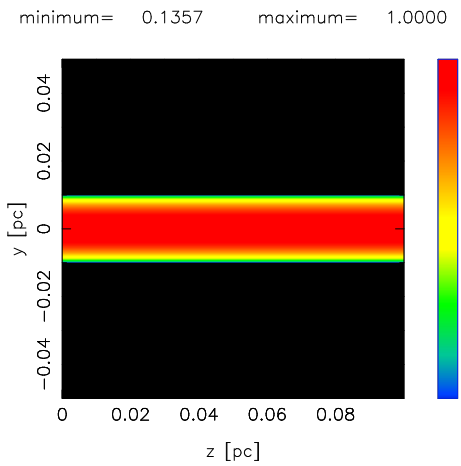}
\caption
{
2D map of the intensity of a jet which has length 0.1~pc
and radius of 0.01pc. 
}
\label{emission_constant}
\end{figure}

A thermal model for the image   is characterised by a
constant temperature and density  in  the internal region of the cylinder.
Therefore, we assume that the number density $C$ is
constant in a cylinder  of radius $a$ and then falls  to 0,
see the simplified transfer equation 
(\ref{transfersimplified}).
 The
line of sight when the observer is situated at the infinity of
the $x$-axis and the cylinder's axis is in the perpendicular position 
is the locus parallel to the $x$-axis, which
crosses  the position $y$ in a Cartesian $x-y$ plane and
terminates at the external circle of radius $a$. 
A similar treatment for the sphere is given in 
\cite{Zaninetti2009a}.
The length  of this locus  in the optically 
thin layer approximation  is 
\begin{eqnarray}
l_{ab} = 2 \times ( \sqrt {a^2 -y^2}) \quad  ;   0 \leq y < a
\quad . 
\label{lengthsphere}
\end{eqnarray}
The number density $C_m$ is constant  in the circle of radius $a$
and therefore the intensity of the radiation is
\begin{eqnarray}
I_{0a} =C_m \times  2 \times ( \sqrt { a^2 -y^2})
 \quad  ;  0 \leq y < a    \quad .
\label{icylinder}
\end{eqnarray}
A typical example of this cut is reported in Figure \ref{cut}
and the intensity of all the cylinder is reported
in Figure \ref{emission_constant}.

\subsection{Numerical image}

\label{secnum} 
\begin{figure}
 \begin{center}
\includegraphics[width=7cm,angle=-90]{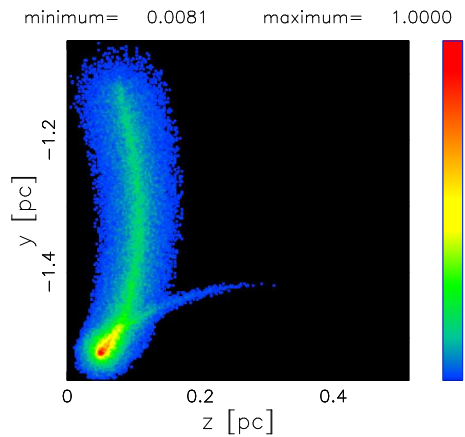}
 \end {center}
\caption {
 2D intensity map  of HH34,
 parameters as in Figure \ref{hh34_traj}.
}
\label{hh34_complex_zoom}
\end{figure}

\begin{figure}
 \begin{center}
\includegraphics[width=7cm]{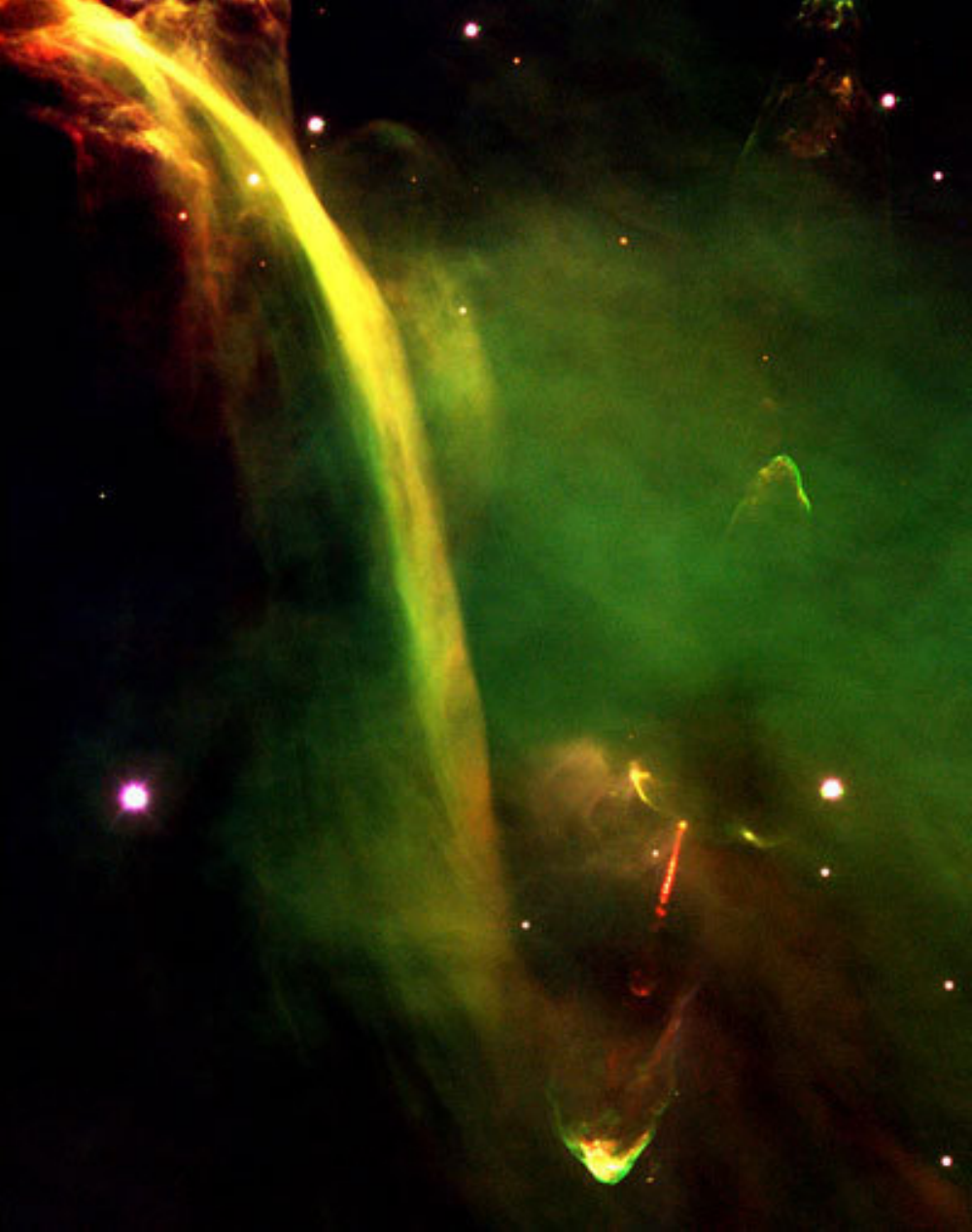}
 \end {center}
\caption {
Three-color composite image of the young object HH34.
}
\label{hh34_vlt}
\end{figure}

The numerical algorithm that allows us to
build  a complex  image 
in the optically thin layer approximation 
is now 
outlined.
\begin{itemize}
\item  
An empty, value=0,
memory grid  ${\mathcal {M}} (i,j,k)$ which  contains
$400^3$ pixels is considered.
\item  
The points which fill the jet in a uniform  way
to simulate the constant density in the emitting particles
are inserted, value =1, in 
${\mathcal {M}} (i,j,k)$ 
\item  Each point of
${\mathcal {M}} (i,j,k)$  has spatial coordinates $x,y,z$ which  can be
represented by the following $1 \times 3$  matrix, $A$,
\begin{equation}
A=
 \left[ \begin {array}{c} x \\\noalign{\medskip}y\\\noalign{\medskip}{
   z}\end {array} \right]
\quad  .
\end{equation}
The orientation  of the object is characterised by
 the
Euler angles $(\Phi, \Theta, \Psi)$
and  therefore  by a total
 $3 \times 3$  rotation matrix,
$E$, see \cite{Goldstein2002}.
The matrix point  is
represented by the following $1 \times 3$  matrix, $B$,
\begin{equation}
B = E \cdot A
\quad .
\end{equation}
\item 
The intensity 2D map is obtained by summing the points of the
rotated images.
\end{itemize}
A typical result of the simulation is  reported
in Figure \ref{hh34_complex_zoom},
which should be compared with the observed image
as given by Figure \ref{hh34_vlt}.

\subsection{The mathematical knots}

\begin{figure}
 \begin{center}
\includegraphics[width=9cm]{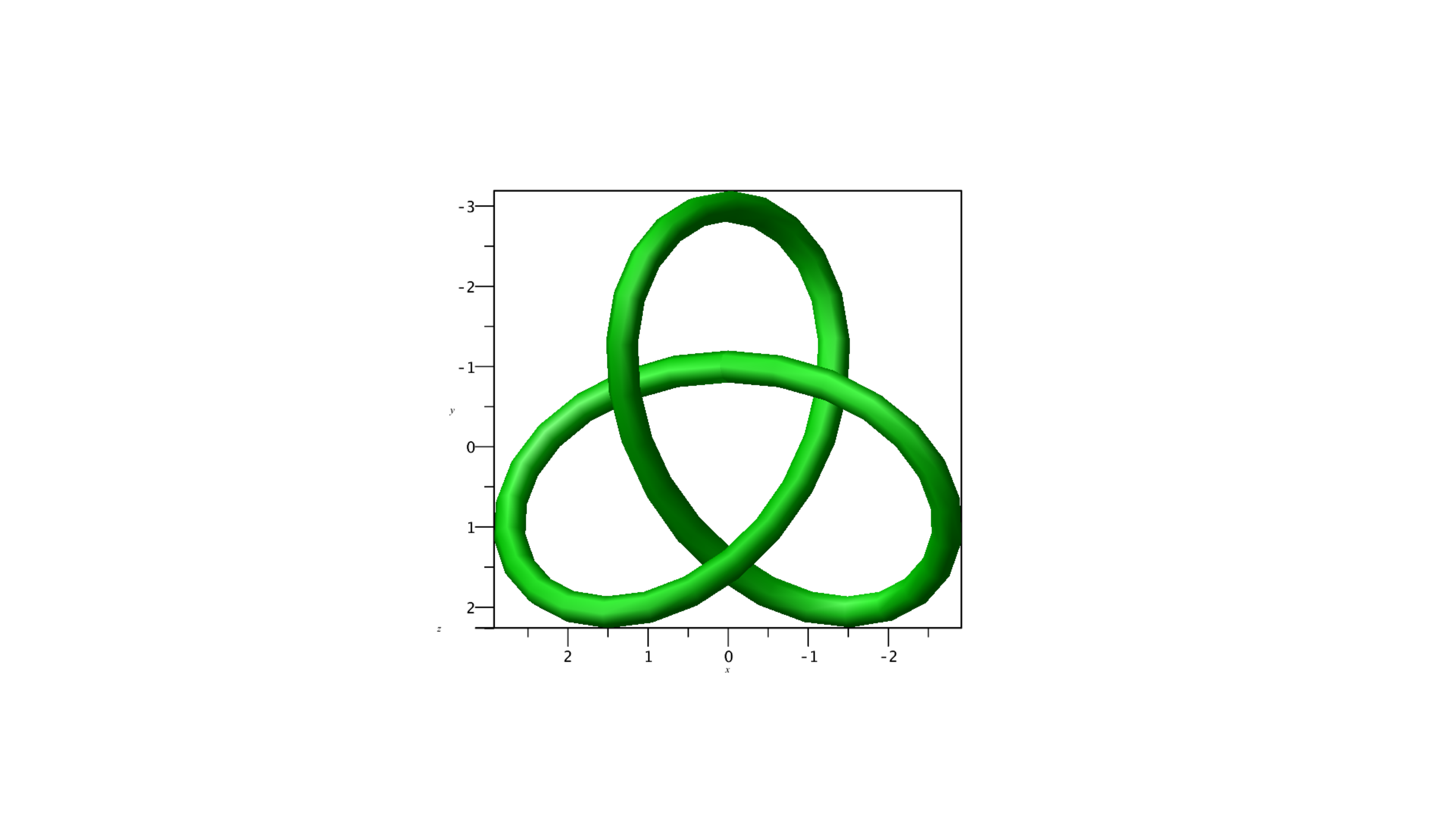}
 \end {center}
\caption 
{
3D view of the trefoil 
when 
the three Eulerian angles 
which characterises  the point of view are 
   $ \Phi   $= 0 $^{\circ }$
 , $ \Theta $= 90 $^{\circ }$
 and $ \Psi $= 0 $^{\circ }$.
}
\label{trefoil}
\end{figure}

The trefoil knot is  defined by the following parametric equations:
\begin{eqnarray}
x =\sin \left( t \right) +2\,\sin \left( 2\,t \right) \\
y =\cos \left( t \right) -2\,\cos \left( 2\,t \right) \\
z =-\sin \left( 3\,t \right)                          
\end{eqnarray}
with $ 0 \leq t \leq 2 *\pi$.
The visual image depends by  the Euler angles,
see Figure \ref{trefoil}.

\begin{figure}
 \begin{center}
\includegraphics[width=9cm,angle=-90]{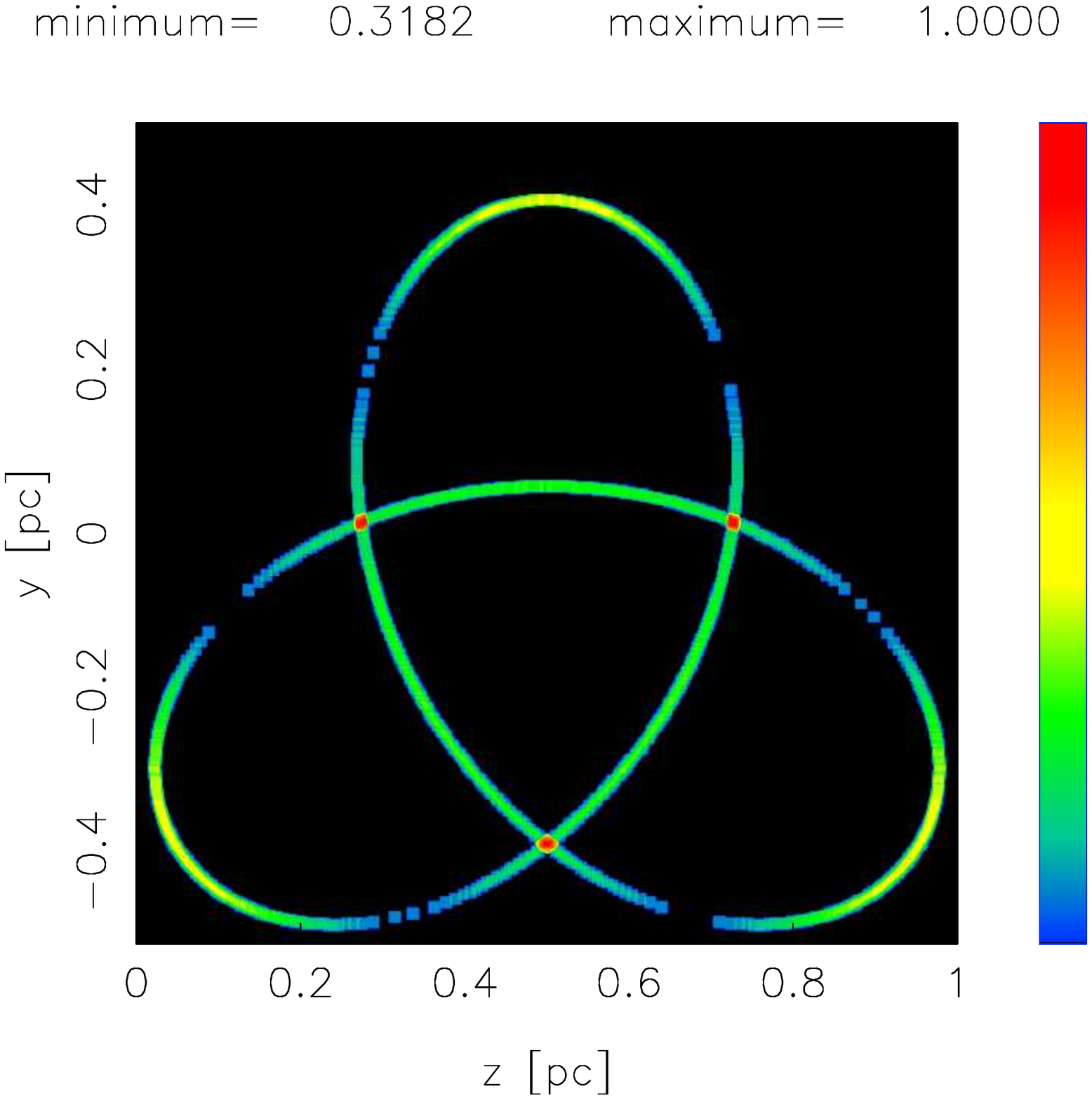}
 \end {center}
\caption 
{
Image of the trefoil with parameters as in Figure \ref{trefoil}, 
the side of the box in pc is 1 and the radius of the
tube in pc is 0.006.
}
\label{emission_knot}
\end{figure}

The image in the optically thin layer approximation 
can be obtained by the numerical method developed 
in Section \ref{secnum} and is reported 
in Figure \ref{emission_knot}.

This 2D map in the theoretical intensity of emission
shows an enhancement   where two mathematical knots
apparently intersect.

\section{Conclusions}

{\bf Laws of motion} 
We analysed two  simple models for the law of motion 
in HH objects 
as given by the Stoke's and Newton's behaviour,
see Section \ref{section_simple}.
A third  law of motion  is used for turbulent jets 
in the presence of a medium whose density decreases with 
a power law, as given by equation \ref{profpower}.
The model that is adopted for the turbulent jets
conserves   the  flux of energy. For example, equation
(\ref{velocitypower}) reports 
the velocity as function of the position.
The $\chi^2$ analysis for observed/theoretical  velocity
as function of time/space, see 
 Table \ref{chi2hh34}, 
assigns the smaller value to the turbulent jet.

{\bf Back reaction}
The insertion  of the back reaction  in the equation of motion 
allows us to introduce 
a finite rather than infinite jet's length, 
see equation (\ref{jetlength}).

{\bf The extended region}

The extended region of HH34 is modeled by combining the 
decreasing jet's velocity  with the constant velocity 
and precession 
of the central object, 
see  final matrix (\ref{eqnmatrix}).

{\bf The theory of the image}

We have analysed the case of an optically thin layer approximation
to provide an explanation for the 
so called "bow shock" that is visible in HH34.
This effect can be reproduced when two  
emitting regions apparently intersect on the plane of the sky,
see the numerical simulation as given by Figure    
\ref{hh34_complex_zoom}.
This   curious effect
of  enhancement in the intensity of emission 
can easily be reproduced  when the image theory is applied 
to the mathematical knots, see the example
of the trefoil in Figure \ref{emission_knot}.

\section*{Acknowledgments}

Credit for Figure  
\ref{hh34_vlt} 
 is  given to  ESO. 

\providecommand{\newblock}{}

\end{document}